\documentclass{PoS}

\usepackage{listings}

\title{Recent developments in the tmLQCD software suite}

\ShortTitle{tmLQCD}

\author{A.~Abdel-Rehim\\
  CaSToRC, The Cyprus Institute, Nicosia, Cyprus\\
  E-mail: \email{a.abdel-Rehim@cyi.ac.cy}}

\author{F.~Burger, B.~Kostrzewa\\
  Humboldt-Universit{\"a}t zu Berlin, Institut f{\"u}r Physik, Berlin, Germany\\
  E-mail: \email{florian.burger@hu-berlin.de,bartosz.kostrzewa@desy.de}}

\author{A.~Deuzeman\\
  Albert Einstein Center for Fundamental Physics - University of Bern, Switzerland\\
  E-mail: \email{albert.deuzeman@gmail.com}}

\author{K.~Jansen\\
  NIC, DESY, Zeuthen, Germany\\
  E-mail: \email{karl.jansen@desy.de}}

\author{L.~Scorzato\\
  Trento Institute for Fundamental Physics and Application (TIFPA), Trento, Italy\\
  E-mail: \email{luigi@scorzato.it}}

\author{\speaker{C.~Urbach}\\
  HISKP (Theory), Rheinische Friedrich-Wilhelms Universit{\"a}t Bonn, Germany\\
  E-mail: \email{urbach@hiskp.uni-bonn.de}}

\abstract{We present an overview of recent developments in the
  tmLQCD software suite. We summarise the features of the
  code, including actions and operators implemented. In particular, we
  discuss the optimisation efforts for modern architectures using
  the Blue Gene/Q system as an example.
  \vspace{0.5cm}

  {\rm\tiny HU-EP-13/59,SFB/CPP-13-84}  
}

\FullConference{31st International Symposium on Lattice Field Theory - LATTICE 2013\\
		July 29 - August 3, 2013\\
		Mainz, Germany}

\begin{document}

\section{Introduction}

The Lattice QCD community relies to a large extent on the efficient
usage of available computer resources. It is, therefore, mandatory to
optimise existing codes for new supercomputer architectures as well as
commodity systems newly appearing on the market.

Due to the increasing complexity and diversity of modern computer
architectures there is need for flexible software which allows one to
quickly implement and test new developments. Furthermore, lattice
QCD actions nowadays simulated have also reached a high level of
diversity, which needs to be mirrored by the software. Writing,
debugging and implementing such software requires, hence, a
non-negligible amount of manpower.

From this perspective it might appear sensible to develop only one,
community wide code basis. However, for scientific hygiene there
should be clearly at least two or three implementations available such
that cross-checks are possible. Ideally, all the different
implementations are publicly available, which increases the chance to
find mistakes in the codes and allows any lattice QCD practitioner to
re-use them (c.f. Ref.~\cite{albert:lat2013}). 

By now there are several lattice QCD software suites available as open
source, among others the MILC code~\cite{milc}, Chroma~\cite{Edwards:2004sx},
openQCD~\cite{openqcd} and bQCD~\cite{Nakamura:2010qh}; another one is
tmLQCD~\cite{Jansen:2009xp} obtained from
github~\cite{tmlqcdgithub}. tmLQCD started as a code for simulations
using the Wilson twisted mass formulation of lattice QCD, but includes by
now a much wider range of actions and lattice Dirac
operators. Moreover, tmLQCD is fully parallelised and includes
optimisations for most modern supercomputer architectures.

\section{tmLQCD: General Overview}

The tmLQCD software is written in the C programming language following
the C99 standard. It ships with an {\ttfamily autoconf} configuration script,
which makes it relatively easy to compile the code on most modern
computer platforms. It also comes with documentation as a \LaTeX\
document. 

Once compiled, tmLQCD offers two executables: firstly an inverter
offering a range of iterative solvers needed for computing
propagators. The second executable implements a Hybrid Monte Carlo (HMC)
algorithm~\cite{Duane:1987de} for generating gauge configurations using Wilson
twisted mass actions, also including the clover term.

The various physical and algorithmic parameters of both programmes can
by chosen by the user using an input file, which has a simple and
human readable syntax. An example of the general section of such an
input file for the HMC might look as follows:
\begin{lstlisting}[frame=single]
L=4                           # spacial lattice extend
T=8                           # time extend
Measurements = 1000           # no. of trajectories
StartCondition = hot 
ReversibilityCheck = yes      # perform reversibility check
ReversibilityCheckIntervall=2 # every second traj.
\end{lstlisting}
The code comes with a selection of sample input files.

The executables can be compiled as scalar or parallel programmes to be
decided at configure time. The parallelisation is 
implemented using a hybrid approach with the Message Passing Interface
(MPI) and openMP. The MPI topology and the number of (openMP) threads
per MPI task can be specified in the input file. 

tmLQCD directly reads and writes the ILDG gauge configuration
format~\cite{DeTar:2007au} and the SCIDAC propagator format 
using the LIME library~\cite{lime}. There is, therefore, full
compatibility to the Chroma software. Moreover, tmLQCD can be
configured to use the Lemon library~\cite{Deuzeman:2011wz}, which is a
parallelised replacement of LIME using the MPI parallel I/O capabilities. Lemon
significantly increases the I/O performance on massively parallel
machines. In particular, the propagator computation, usually heavily
I/O bound, benefits from the usage of Lemon.

\section{Iterative Solvers}

One of the main tasks in lattice QCD is solving
\begin{equation}
  \label{eq:dirac}
  D\ \cdot\ \psi\ =\ \eta
\end{equation}
for $\psi$, where $D$ is some discretisation of the gauge covariant
Dirac operator. Note that we have suppressed all indices for
simplicity. The lattice Dirac operator can be viewed as a large sparse
matrix, which makes iterative solvers and in particular Krylov space
solvers like the conjugate gradient (CG) most suited for solving
equation \ref{eq:dirac}, see Ref.~\cite{saad:2003a} for a general
discussion. 

Several discretisations of the Dirac operator are implemented in
tmLQCD: the Wilson and Wilson twisted mass Dirac operators both with
and without clover term, the non-degenerate Wilson twisted mass Dirac
operator~\cite{Frezzotti:2000nk} with and without clover term and the overlap
operator~\cite{Neuberger:1997fp,Neuberger:1998wv}. For Wilson type
operators also even/odd 
preconditioning~\cite{DeGrand:1990dk} is implemented. The clover operators are
currently only available with even/odd preconditioning. The operator
can be specified in the input file like in the following example for a
(mass degenerate) even/odd preconditioned Wilson twisted mass Dirac
operator:
\begin{lstlisting}[frame=single]
BeginOperator TMWILSON
  2kappaMu = 0.05
  kappa = 0.177
  UseEvenOdd = yes
  Solver = CG
  SolverPrecision = 1e-14
  MaxSolverIterations = 1000
EndOperator
\end{lstlisting}
The physical parameters are the $\kappa=0.177$ hopping parameter and
the twisted mass parameter $\mu$, the latter specified as
$2\kappa\mu=0.05$. In the above listing the iterative solver to be
used is CG, the solver precision is set to $\|D\psi - \eta\|^2\leq
10^{-14}$ and not more than $1000$ iterations should be used.

The list of available solvers includes CG, BiCG, BiCGstab, FGMRES,
CGS, EigCG~\cite{Stathopoulos:2007zi} and GCR. A FGMRES solver applying inexact deflation as discussed in
Ref.~\cite{Luscher:2007se} is also available, as well as a multiple mass CG
solver for the twisted mass Dirac operator~\cite{Chiarappa:2006hz}. Note that the
optimal (Krylov) solver depends on the particular discretisation of
the lattice Dirac operator.

\section{Lattice Actions}

Lattice QCD actions are usually split into a sum of a gauge and a
fermionic part. Concerning the gauge part, tmLQCD implements the
Wilson plaquette gauge action and the family of gauge actions
including an additional planar $2\times1$ rectangular Wilson
loop. These include the tree level Symanzik improved, the Iwasaki and
the DBW2 gauge actions.

For the fermionic part we have implemented several so-called
pseudofermion actions. They are based on the stochastic representation
of a determinant of a matrix $Q^2$
\[
\det(Q^2) = \int\mathcal{D}\phi^\dagger\, \mathcal{D}\phi\ e^{-\phi^\dagger\frac{1}{Q^2}\phi}\,,
\]
where the pseudofermion fields $\phi^\dagger,\phi$ follow bosonic
statistics. Following the notation of Chroma, we call one such term a
monomial and an example for an input file is as follows:
\begin{lstlisting}[frame=single]
BeginMonomial DET
  Timescale = 1              # time scale to integrate on
  2KappaMu = 0.177
  kappa = 0.177
  Solver = CG
  AcceptancePrecision =1e-20 # accept/reject precision
  ForcePrecision = 1e-12     # MD evolution precision
EndMonomial
\end{lstlisting}
It corresponds to a two flavour, mass degenerate Wilson twisted mass
Dirac operator pseudofermion monomial. You may specify the solver used
in the HMC update, as well as the precisions for the molecular
dynamics (MD) evolution and the accept/reject step.

There is a list of other monomials supported: for two mass degenerate
Wilson and Wilson clover twisted mass fermions there are besides the
``$\det$'' monomials also ratios of determinants needed for Hasenbusch
mass preconditioning with multiple
timescales~\cite{Hasenbusch:2001ne,Hasenbusch:2002ai,Urbach:2005ji}.
For the mass non-degenerate Wilson 
and Wilson clover twisted mass doublet there is a
polynomial~\cite{Frezzotti:1998eu} and a rational 
monomial~\cite{Clark:2006fx} implemented. Finally, for a single Wilson clover fermion a
rational monomial can be used. 

tmLQCD offers different schemes for integrating the MD
equations of motion: the simple leap-frog integration scheme, the
second order minimal norm scheme and a fourth order Omelyan
scheme~\cite{Sexton:1992nu,Omelyan:2003:SAI}. They can be combined on different time scales, as can be seen
from the following input file listing:
\begin{lstlisting}[frame=single]
BeginIntegrator 
  Type0 = LEAPFROG       # integrator on timescale 0
  Type1 = 2MN            # integrator on timescale 1
  IntegrationSteps0 = 1
  IntegrationSteps1 = 2
  Tau = 1                # trajectory length
  NumberOfTimescales = 2
EndIntegrator
\end{lstlisting}
The zeroth time scale of the two is the finest, the total trajectory
length is specified using $\tau=1$ and the number of steps for time
scale {\ttfamily N} by {\ttfamily IntegrationStepsN}. The
step numbers are defined recursively. Therefore, the step length on
timescale zero is given by
$\Delta\tau_0=\tau/N_1/N_0=\Delta\tau_1/N_0$, where $N_i$ is 
the number of steps on timescale $i$. Note that there are additional
factors of $1/2$ for the higher order schemes.

\section{Optimisation: Example BG/Q}

As mentioned before, tmLQCD includes optimisations for several modern
supercomputer architectures, IBMs Blue Gene/Q, Intels SSE instruction
set and the Aurora architecture. We also have an inverter and parts of
the HMC implemented for NVIDIA GPUs.

Here we will discuss as an example the BG/Q architecture. The
BG/Q compute nodes consist of one CPU with 16 cores with
four hardware threads each. Hence, in total there
are 64 hardware threads per node, which can be divided into MPI tasks
and/or (openMP) threads. The nodes are connected via a five
dimensional torus network. For a first discussion on how to port
lattice QCD codes for BG/Q see Ref.~\cite{Boyle:2012iy}. The floating
point unit (FPU) includes a four double wide SIMD vector unit (QPX). For
maximal performance it is mandatory to utilise it appropriately. In
total one node of the BG/Q has a peak performance of $204.8$ Gflop/s.

\begin{figure}[t]
  \centering
  \includegraphics[width=.65\linewidth]{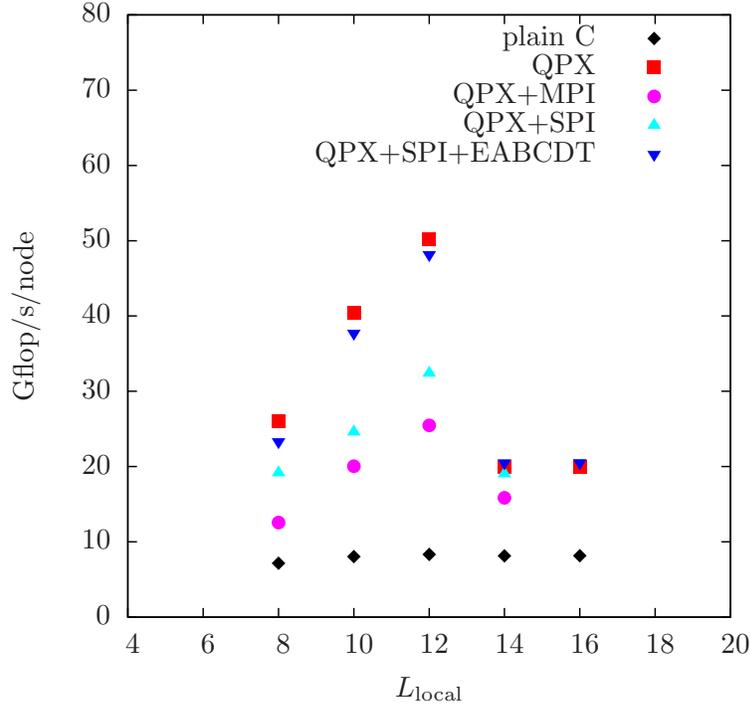}
  \caption{Single BG/Q node double precision performance of the
    hopping matrix in Gflop/s as a 
    function of $L_\mathrm{local}$. ``Plain C'' and ``QPX'' correspond
    to measurements with communication switched off using a plain C 
    implementation and one including QPX instructions, respectively. The
    other points include communication, see text for details.}
  \label{fig:bgq}
\end{figure}

For the following discussion we used a hybrid MPI/openMP
implementation with always $64$ openMP threads and one MPI task per
BG/Q node. For a more detailed discussion on how to optimally use
openMP see the contribution~\cite{bartekcode:lat2013} at this conference.
In figure~\ref{fig:bgq} we show the performance of the
tmLQCD Wilson Dirac hopping matrix in Gflop/s per BG/Q node as a
function of the node local lattice extent $L_\mathrm{local}$. Each
node worked on a local lattice volume of $L_\mathrm{local}^4$. 

First we investigated the code performance {\it without
internode communication}. The plain C99 implementation is shown as black
diamonds leading to less than 5\% of peak performance, almost
independently of the local volume. This result points towards a badly
saturated FPU. The red squares represent the code including the QPX
instruction set utilising the intrinsic functions provided by the IBM
C compiler. A strong improvement is visible, with up to 25\% of peak at
$L_\mathrm{local}=12$. For $L_\mathrm{local}\geq14$ the local problem
does no longer fit into the cache leading to a plateau in the
performance around $20$ Gflop/s per node.

The purple circles in figure~\ref{fig:bgq} represent the hopping
matrix with QPX instructions, but now {\it with internode
  communication switched on}. The MPI overhead turns out to be
significant, more than halving the performance where the local volume
fits into cache. 

Circumventing this problem is possible by overlapping communication
and computation. For this purpose we use an implementation of the
hopping matrix where in a first step we project to half size spinors
$\phi^+$, $\phi^-$ for all $x$ and $\mu$
\[
\phi^+(x-\hat\mu, \mu) =
U_\mu(x-\hat\mu)P_\mu^{(\textrm{full}\rightarrow\textrm{half})}(1-\gamma_\mu)\psi(x)\,,\
\phi^-(x+\hat\mu,\mu) =
P_\mu^{(\textrm{full}\rightarrow\textrm{half})}(1+\gamma_\mu)\psi(x)\,.  
\]
In a second step the communication is performed and finally the result
is generated by reconstructing the full spinor for all $x$
\[
\eta(x) = \sum_\mu\left[ P_\mu^{(\textrm{half}\rightarrow\textrm{full})}\phi^+(x,\mu) +
 P_\mu^{(\textrm{half}\rightarrow\textrm{full})}U_\mu^\dagger(x-\hat\mu)\phi^-(x,\mu)\right]\,.
\]
The field $\phi^{+,-}$ requires only half the amount of data to be
communicated compared to the full spinor field $\psi$. Overlapping
communication and computation can now be achieved by dividing all $x$
into one set $x_\mathrm{surface}$ residing on the local surface and a
second set $x_\mathrm{bulk}$. Then the projection to $\phi^{+,-}$ is
first done for all $x_\mathrm{surface}$, then the communication is started
non-blocking in parallel to the the computation for
$x_\mathrm{bulk}$. When the latter computation is finished it is
checked whether the communication is finished also (c.f. also
\cite{Brambilla:2012jc}).

It turns out that this does not work as expected using the MPI
library provided on the BG/Q. All the communication is effectively
done in the {\ttfamily MPI\_Wait} and, therefore, there is no
gain. However, IBM provides a lower level communication library called
SPI. Replacing the MPI calls with SPI calls leads to the light blue
triangles in figure~\ref{fig:bgq} and, thus, to a significant
improvement. The blue inverted triangles finally include also the correct
mapping of the lattice QCD geometry to the five dimensional BG/Q
network. So, eventually we obtain almost the same performance as in
the case with communication switched off.

\section*{Acknowledgements}

We thank all members of ETMC for the most enjoyable collaboration. 
B.K. is supported by the National Research Fund, Luxembourg. This work
is supported in part by DFG and NSFC (CRC 110) and by DFG
SFB/TR9. A. A.-R. acknowledges support from the PRACE-2IP project
under grant number EC-RI-283493. LS thanks the SUMA project for partial support.

\bibliographystyle{h-physrev5}
\bibliography{bibliography}

\end{document}